\documentclass[prl,twocolumn,epsf,psfig]{revtex4}
\usepackage{graphicx}
\DeclareGraphicsExtensions{.pdf,.png,.gif,.jpg}
\usepackage{float}
\usepackage{subfig}
\usepackage{epsfig}
\usepackage{wrapfig}
\usepackage{bbold}
\usepackage{amsmath}
\restylefloat{figure}

\begin{document}

\title{Massive electrons and unconventional room-temperature superconductivity in superhydrides}

\author{Theja N. De Silva}
\affiliation{Department of Chemistry and Physics,
Augusta University, Augusta, Georgia 30912, USA.}

\begin{abstract}
The search for room-temperature superconducting materials has been at the center of modern research for decades. The recent discovery of high-temperature superconductivity, under extreme pressure in hydrogen-rich materials, is a tremendous achievement in this research front. This discovery offers a route in the search for room temperature superconductivity at ambient pressure. The superconductivity of these hydrogen-rich materials was confirmed by the observation of zero-resistance, isotope effects, effect of magnetic field, and other standard properties. However, some of the experimental features were puzzling as they were not consistent with the known superconductivity theories. These debatable features have lead to a series of recent publications downplaying the existence of superconductivity in these superhydrides. Here we propose a concept of massive electrons under pressure and successfully explain all non-standard experimental observations. Our massive electron concept explains the large effective mass of the quasiparticles, the reason for the high critical temperatures for moderate electron-phonon couplings, and a 3-5 orders of magnitude larger conductivity causing a narrow resistivity broadening at the transition in the presence of magnetic field. We anticipate our findings will lead to a new directions and tweaks in current research in the search for ambient-pressure, room-temperature superconductors.
\end{abstract}

\maketitle

\section{I. Introduction}

Superconductivity research has been at the heart of modern condensed matter physics and material science research for decades. As superconductors can conduct electric current without resistance, the potential application of superconductors in technology can have a revolutionized impact. There are two types of superconductors that have been discovered to date: conventional and unconventional superconductors. Conventional superconductors are understood as materials whose superconductivity originates from electron-phonon interactions. The properties of these phonon-driven superconductors can be described by the Bardeen-Cooper-Schrieffer (BCS) and Migdal-Eliashberg theories~\cite{bcs}. The superconductivity of unconventional superconductors is believed to be driven by strong electron-electron correlations. The most prominent unconventional superconductors are cuprates~\cite{cuoA, cuoB}, iron pnictides~\cite{ironpA, ironpB}, and nickalates~\cite{nickalateA, nickalateB, nickalateC}. Even though the nature of superconductivity in unconventional materials is not fully understood, it is known that both of these superconducting types exhibit standard properties. These properties include the Meissner effect, upper and lower critical magnetic fields, critical currents, and resistive transition at critical temperatures.

The longstanding challenge in superconductivity research was finding a room temperature superconducting material. The search for room temperature superconductivity was renewed after the discovery of unconventional copper based superconductors with critical temperatures as high as 133 K at ambient pressure. The BCS theory provides clues for achieving high critical temperatures for conventional superconductors. The theory suggests that high frequency Debye phonons, strong electron-phonon interactions, and high density of states can enhance the critical temperatures of conventional superconductors. Following these clues, magnesium diboride (MgB$_2$) has been synthesized and found to be superconducting below 39 K at ambient pressure~\cite{mgb2A, mgb2B}. The high frequency phonon spectrum due to the light elements in MgB$_2$ is believed to be the reason for this relatively higher critical temperature. Based on the idea of high phonon frequency due to the light hydrogen atom, Ashcroft proposed the possibility of having high temperature phonon based superconductivity in hydrogen rich materials, if attractive pairing interaction exists~\cite{ashcroft1, ashcroft2}. Ascroft’s proposal was pioneered by the idea of a \emph{chemical pre-compression} effect proposed by Gilman~\cite{gilman}. Gilman’s proposal of achieving high temperature superconductors came soon after the discovery of ambient pressure hydrogen rich superconductors Th$_{4}$H$_{15}$ at a critical temperature of 8 K~\cite{THsc}. Motivated by Gilman’s idea and the discovery of a hydride superconductor, subsequent studies on Pd–H and Pd–Cu–H systems were reported to exhibit superconductivity below 10 K~\cite{pdH}. Despite the support of calculations showing that metallic hydrogen is a good candidate for a room temperature superconductor, all experimental efforts turned out be negative for pure hydrogen. Therefore, researchers shifted their efforts toward binary and ternary hydride compounds. The density functional theory, Monte-Carlo, and other numerically based calculations~\cite{numcA, numcB, numcC,numcD, numcE, numcF, numcG, numcH, numcI, numcJ, numcK} and the discovery of phonon based high temperature superconductivity in H$_3$S at high pressure~\cite{SCEX1} ignited a wealth of research by synthesizing superhydrides at very high pressure values. To date, about dozen synthesized superhydrides have shown to be near-room temperature superconductors at high pressure. These include phosphorous hydrides~\cite{exp1}, lanthanum hydrides~\cite{exp2, exp3, exp4}, yttrium hydrides~\cite{exp5, exp6, exp7}, thorium hydrides~\cite{exp8}, binary cerium hydrides~\cite{exp9}, ternary lanthanum-yttrium hydrides~\cite{exp10} and carbonaceous-sulfur hydride ternary compounds~\cite{SCEX2}. The most notable among these compounds are the lanthanum hydride~\cite{exp2, exp3, exp4} and carbonaceous sulfur hydride (C-S-H) systems~\cite{SCEX2}. Two recent experiments by Drozdov \emph{et al}~\cite{exp2}  and Snider \emph{et al}~\cite{SCEX2} report near-room temperature superconductivity for LaH$_{10}$ at pressure 267 GPa and room temperature superconductivity for C-S-H at pressure 275 GPa. The superconductivity of these compounds was confirmed by the observation of zero resistance and magnetic susceptibility. Furthermore, these experiments show a decrease in critical temperature in the presence of an external magnetic field. The conventional nature of the superconductivity in these compounds was confirmed by a pronounced isotope effect on the critical temperature. The experimental estimates further support that these clathrate-like hydrides superconductors are strongly type-II.

For both conventional and unconventional, the magnetic field responses to type-I and type-II superconductors are very different. The type-I superconductors completely expel the magnetic field up to the T-temperature dependent critical critical field $H_c(T)$, beyond at which it becomes normal metal. This perfect diamagnetism can be described by a supercurrent circulating within a thin surface layer of the superconductor. The thickness of this surface layer is a temperature dependent material parameter known as the London penetration depth $\lambda(T)$. On the other hand, the type-II superconductors are perfect diamagnet only up to the lower critical field $H_{c1}(T) < H_c(T)$. In the range of magnetic field up to the upper critical field $H_{c1}(T) < H < H_{c2}(T)$, the magnetic flux can penetrate into the material in the form of vortices and can give rise flow resistance to the critical current. The Ginzburg-Landau parameter $\kappa = \lambda(0)/\xi(0)$ which is approximately a temperature independent quantity can be used to determine whether a superconductor is type-I or type-II. Here, the coherence length $\xi(T) = (\xi_0^{-1} + l^{-1})^{-1}$ is usually taken as the shortest of the Pippard coherence length $\xi_0(T)$ or the mean free path $l(T)$. All experimental evidence suggests that superhydride superconductors are type-II as $\kappa \gg 1$.

Most of the experimental data strongly supports the existence of superconductivity in superhydrides. However, some of the experimental features are puzzling and seem to violate known standard superconducting properties. While some features support type-I superconductivity, others support type-II. This disparity shows the simultaneous coexistence of type-I and type-II. Another puzzling question is the high critical temperature with moderate electron-phonon coupling. Theoretical calculations suggest that the dimensionless electron-phonon coupling $\Lambda \sim 2$ in superhydrides have moderate values~\cite{numcD}. In general, the width of the resistive transition in the presence of a magnetic field is expected to be large for type-II superconductors. For example, the width of the resistive transition in MgB$_2$ at a magnetic field $H = 0.15 H_{c2}$ is about $\Delta T_C/T_C \sim 0.15 \%$. In contrast, the resistive transition width of C-S-H at the same magnetic field is smaller than that of MgB$_2$ by a factor of about 50~\cite{mgb2B, hirsch1}. This narrow transition width in the C-S-H system and other  superhydrides apparently suggests that these are type-I superconductors~\cite{SCEX2}. Further, using an experimental sample size and measured resistance, the resistivity of the C-S-H system was calculated by Dogan \emph{et al}~\cite{dogan}. The calculation was done using a resistive formula derived from the four-point van der Pauw procedure. The calculated resistivity was found to fall into the poor metal/semimetal range above the critical temperature. The resistivity below the critical temperature was found to be 2-3 orders of magnitude lower falling into the typical metal range. In addition, by analyzing resistivity broadening experimental data, Hirsch \emph{et al}~\cite{hirsch2} argued that the zero-temperature critical current density in C-S-H systems is five orders of magnitude larger than that of standard superconductors. The experimental data reported in Ref.~\cite{SCEX1} for the H$_3$S system indicates that the effective mass of the electrons at pressure 150 GPa is larger than the expected effective mass by a factor of about 10~\cite{hirsch3}. This mass enhancement is not consistent with the electron-phonon interactions estimated for H$_3$S, nor the theoretical calculations~\cite{numcA, numcB, numcC, numcD, numcE, numcF, numcG}. Due to the fact that these experimental features are not able to be explained using the standard superconductivity theories, a series of recent articles argue that the superhydrides under pressure are either a unique kind of superconductors or not superconductors at all~\cite{hirsch1, dogan, hirsch2, hirsch3, hirsch4}.

In this paper, we successfully answer all debatable experimental observations above using a massive electron concept. We show that the effective mass of the electrons exponentially increases with pressure. The mass enhancement makes the density of states larger resulting in strong effective interactions between electrons at high pressure. Thus, the superhydrides under pressure are strongly interacting conventional BCS superconductors. However the conventional classification of type-I versus type-II is not applicable to the superhydrides as the coherence length and the penetration depth are pressure dependent. We show that the narrow width of the resistivity transition originates from the flow resistivity of vortices in the presence of a magnetic field. We find that the flux flow resistivity is exponentially smaller at high pressure due to the pressure dependence on the coherence length.

\section{II. Pressure dependence on the effective mass}

In this section, we briefly illustrate the pressure dependence on the effective mass using a simplified picture. Let's consider the pressure change on the material unit cell $\Delta P \equiv P = P_{ex} - P_0$, where $P_{ex}$ and $P_0$ are the applied pressure and the ambient pressure, respectively. The volume of the unit cell shrinks under the applied pressure so the change in volume can be written as,

\begin{eqnarray}
\frac{V - V_0}{V_0} = - K_V  \Delta P,
\end{eqnarray}

\noindent where $K_V$ is the compressibility and $V - V_0$ is the change in volume. The onsite Coulomb repulsion $U(P)$ between electrons increases as the cell volume decreases,

\begin{eqnarray}
U(P_{ex}) -U(P_0) = -K_U \frac{V - V_0}{V_0} \\ \nonumber
\frac{U(P_{ex}) -U(P_0)}{U(P_0)} = K_C dP,
\end{eqnarray}

\noindent where $K_C = K_UK_V/U(P_0)$ is a material dependent constant. Approximating $U(P_0) \rightarrow U(P)$, we find,

\begin{eqnarray}
\frac{1}{U} \frac{dU}{dP} = K_C.
\end{eqnarray}

\noindent The pressure dependence of the onsite repulsion is then given by the solution of this equation, $U = U_0 e^{K_CP}$. The pressure dependence on the tunneling energy or the hopping integral $t$ can also be approximated in a similar fashion. The pressure dependence on the tunneling energy then has the form $t = t_0 e^{-K_t P}$.

The electronic part of the effective Hamiltonian for the propagation of quasiparticles can be written as,

\begin{eqnarray}
H_0 = \sum_k \epsilon_k c^\dagger_{k \sigma} c_{k \sigma},
\end{eqnarray}

\noindent where $c^\dagger_{k \sigma}/c_{k \sigma}$ represents the creation/annihilation of a quasiparticle of wavevector $k$ with spin $\sigma = \uparrow, \downarrow$. Regardless of the lattice structure, the energy dispersion of the weakly interacting quasiparticles has the form $\epsilon_k = -2 t \sum_\delta \cos (\vec{k} \cdot \vec{\delta})$, where $\vec{\delta}$ is the nearest neighbor lattice vector. For the case of strongly interacting electronic systems, one can consider propagation of holes in the presence of doping in the background of anti-ferromagnetism~\cite{efm1, efm2}. In this case, the quasiparticle dispersion has the form $\epsilon_k = -(2 t^2/U) \sum_\delta \cos (\vec{k} \cdot \vec{\delta})$. In the continuity limit, the quasiparticle dispersion can be approximated by expanding the cosine term to get $\epsilon \sim \hbar^2 k^2/(2 m^{\ast})$, where $m^{\ast}$ is the effective mass of the quasiparticles and $\hbar$ is the Planck's constant. The effective mass of the quasiparticle is $m^{\ast} = \hbar^2/(2 \delta ta_0)$ and $m^{\ast} = \hbar^2U/(2 \delta t^2a_0)$ for the weakly interacting electrons systems and strongly interacting holes systems, respectively. Here $a_0$ is the lattice constant of the underlying host lattice. Using the pressure dependence of the interaction parameters presented before, the effective mass of the relevant quasiparticles responsible for superconductivity in superhydrides under pressure can be written in the form,

\begin{eqnarray}
m^{\ast} = m_0 e^{KP},
\end{eqnarray}

\noindent where $m_0 = m_e(1 + \Lambda)$ with bare electron mass $m_e$ and dimensionless electron-phonon coupling $\Lambda$. Notice, here $P$ is the pressure relative to the ambient pressure, $K$ is a material dependent parameter, and we have neglected the pressure dependence on $\Lambda$. As we see in the following sections, neglecting pressure dependence on $\Lambda$ has no effect on our conclusions. The material dependent parameter $K$ encapsulates the structural and lattice details of the system.

\section{III. Determination of the material dependent parameter $K$}

We start with the reduced BCS Hamiltonian in the mean-field approximation,

\begin{eqnarray}
H = \sum_{k, \sigma} \xi_k c^\dagger_{k \sigma} c_{k \sigma} + \sum_k (\Delta_k c^\dagger_{k \uparrow} c^\dagger_{-k \downarrow} + \Delta_k^\ast c_{-k \downarrow} c_{k \uparrow}),
\end{eqnarray}

\noindent where $\Delta_k = \sum_{k^\prime} V_{k, k^\prime} \langle c_{-k^\prime \downarrow} c_{k^\prime \uparrow} \rangle$ is the superconducting order parameter defined through the thermal expectation value $\langle c_{-k^\prime \downarrow} c_{k^\prime \uparrow} \rangle$ with respect to the Hamiltonian $H$. The momentum conserving effective attractive interaction between quasiparticles $V_{k, k^\prime}$ originates from the electron-phonon interaction. Notice that we are working in the grand canonical ensemble to take care of the conservation of quasiparticle number, so we defined $\xi_k = \epsilon_k - \mu$, where $\mu$ is the chemical potential. The diagonalization of the Hamiltonian is straight forward using the usual Bogoliubov transformation,

\begin{eqnarray}
c_{k \sigma} = \cos (\theta_k) \gamma_k - \sigma \sin (\theta_k) e^{i \phi_k} \gamma^\dagger_{-k, -\sigma},
\end{eqnarray}

\noindent to get,

\begin{eqnarray}
H = \sum_{k \sigma} E_k \gamma^\dagger_{k \sigma} \gamma_{k \sigma} + \sum_k(\xi_k - E_k),
\end{eqnarray}

\noindent where the superconducting energy gap $E_k = \sqrt{\xi_k^2 + \Delta_k^2}$ and the coherence factor is

\begin{eqnarray}
\cos \theta_k = \sqrt{\frac{E_k + \xi_k}{2E_k}}.
\end{eqnarray}

Deriving the thermal expectation value $\langle c_{-k^\prime \downarrow} c_{k^\prime \uparrow} \rangle$ with respect to the diagonalized Hamiltonian and relating it to the superconducting order parameter, the finite temperature gap equation has the form,

\begin{eqnarray}
\Delta_k = -\sum_{k^\prime} V_{k, k^\prime} \frac{\Delta_{k^\prime}}{2 E_k^\prime} \tanh \biggr(\frac{E_{k^\prime}}{2k_BT} \biggr).
\end{eqnarray}

\noindent Following the traditional BCS formalism, we take the interaction to be attractive, $V_{k, k^\prime} = -v/V < 0$ only for when $\xi_k$ and $\xi_{k^\prime}$ are within an energy $\hbar \omega_D$ and zero otherwise. For phonon-mediated superhydride superconductors, $\omega_D$ is the phonon bandwidth known as Debye frequency. This form of the interaction allows us to take $\Delta_k = \Delta e^{i \phi}$ for $|\xi_k| < \hbar \omega_D$, and $\Delta_k = 0$ otherwise. Assuming the density of states for both spins $g(\epsilon)$ is slowly varying in the vicinity of chemical potential $\mu \simeq \epsilon_F$, we have,

\begin{eqnarray}
1 = \frac{g(\epsilon_F) v}{2} \int_0^{\hbar \omega_D} \frac{d\xi}{\sqrt{\xi^2 + \Delta^2}} \tanh \biggr( \frac{\sqrt{\xi^2 +\Delta^2}}{2 k_BT} \biggr).
\end{eqnarray}

\noindent The density of states at the Fermi energy $\epsilon_F$ can be written as $g(\epsilon_F) = g_0(\epsilon_F) e^{KP}$, where the density of states at ambient pressure is

\begin{eqnarray}
g_0(\epsilon_F) = \biggr(\frac{3 n}{\pi^4 \hbar^6}  \biggr)^{1/3} m_0.
\end{eqnarray}

\noindent The quasiparticle density $n$ is related to the Fermi wavevector $k_F$ as $k_F = (3 \pi^2 n)^{1/3}$. The gap equation can be used to solve for the zero temperature order parameter $\Delta_0 \equiv \Delta (T = 0)$,

\begin{eqnarray}
\Delta_0 = \frac{\hbar \omega_D}{\sinh\biggr( \frac{2}{g(\epsilon_F) v}\biggr)}.
\end{eqnarray}

\noindent The finite temperature gap equation at the critical temperature $T = T_C$ can be used to determine the critical temperature of the system. Setting $\Delta(T = T_C) = 0$ and changing the variable $s = \xi/(2k_B T_c)$, we have,

\begin{eqnarray}
\frac{2}{g(\epsilon_F) v} = \int_0^{S_0/2} \frac{\tanh(s)}{s} ds,
\end{eqnarray}

\noindent where $s_0 = \hbar \omega_d/k_BT_C$. By approximating $tanh(s)/s \simeq (1 +s^2)^{-1/2}$ and completing the integration, we find the critical temperature,

\begin{eqnarray}
k_BT_C = \frac{\hbar \omega_D}{2} \frac{1}{\sinh[2/g(\epsilon_F) v]}.
\end{eqnarray}

\noindent The critical temperature of the weak coupling superconductors $T_C(w)$ can be estimated by the large $x = 2/g(\epsilon_F) v$ expansion of the function $[\sinh (x)]^{-1} \rightarrow 2 e^{-x}$, where we find
\begin{eqnarray}
k_BT_C(w) = \hbar \omega_D e^{-\frac{2}{g(\epsilon_F) v}}.
\end{eqnarray}

\noindent This is only a factor of $2 e^{C}/\pi = 1.134$ smaller than the well established critical temperature of weak coupling BCS superconductors, where $C = 0.577215$ is the Euler-Mascheroni constant~\cite{tinkham96}.

The superhydrides under pressure are strong coupling superconductors due to the large effective interaction parameter $g(\epsilon_F) v$. This is due to the large density of states entering through the effective mass. Therefore, we find the critical temperature of the strong coupling superhydrides superconductors at higher pressure values using the small $x$ expansion of the function $\sinh (x) \rightarrow x$:

\begin{eqnarray}
k_BT_C = \frac{\hbar \omega_D v g_0(\epsilon_F) }{4} e^{KP}.
\end{eqnarray}

\begin{figure}[h!]
\includegraphics[width=\columnwidth]{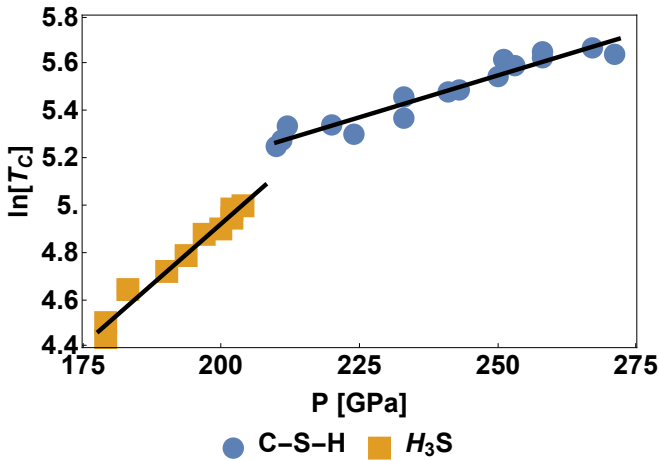}
\caption{(color online) Linear pressure dependence on $\ln{T_C}$ at high pressure, where $T_C$ is the critical temperature. The orange squares represent the experimental data for H$_3$S system extracted from FIG. 1 of Ref. ~\cite{SCEX1}. The blue circles represent the experimental data for carbonaceous  sulfur  hydride (C-S-H) presented in Ref.~\cite{SCEX2}. The solid lines are the linear fit for experimental data.}\label{fit}
\end{figure}

\noindent This clearly shows that the $\ln (T_C)$ has a linear dependence on the pressure at high pressure values and the slope of the $\ln (T_C)$ vs $P$ is the material dependent parameter $K$. We find the $K$ values for both C-H-S and H$_3$S systems using the experimental values of critical temperature. As shown in FIG. \ref{fit}, the experimental data has a clear linear dependence on the pressure, indicating the validity of our theory. Using a linear fit to the experimental data, we find the $K$ values for the C-S-H system and H$_3$S system, $K_{CHS} = 0.007/$GPa and $K_{HS} = 0.021/$ GPa, respectively. See FIG.~\ref{fit} for details.

\section{IV. Pressure dependence on the coherence length, the London penetration depth, and the Ginzburg-Landau parameter}

Let's start with the standard BCS formulas for the coherence length $\xi(T, P)$, and the London penetration depth $\lambda_{L}(T, P)$~\cite{tinkham96}, where we include the arguments of pressure $P = P_{ext} - P_0$ dependence in the definitions.

\begin{eqnarray}
\xi(T, P) =\frac{\hbar \nu_F}{\pi \Delta},
\end{eqnarray}

\noindent where the Fermi velocity $\nu_F = k_F/m^{\ast}$. The Fermi wavevector is related to the density of quasiparticles $n$ through $k_F = (3 \pi^2 n)^{1/3}$. The London penetration depth,

\begin{eqnarray}
\lambda_{L}(T, P) =\biggr(\frac{m^{\ast} c^2}{4 \pi n e^2}\biggr)^{1/2},
\end{eqnarray}

\noindent can be written in terms of the coherence length and the explicit mass dependence,

\begin{eqnarray}
\lambda_{L}(T, P) =\biggr(\frac{3 \hbar^2 c^2}{4 \pi^2 e^2 m_0^2}\biggr)^{1/2} \biggr(\frac{1}{\Delta^3 \xi(T, P)^3}\biggr)^{1/2} \frac{m_0}{m^{\ast}},
\end{eqnarray}

\noindent where $c$ is the speed of light and $e$ is the electron charge. For the purpose of comparison, we provide the zero-temperature London penetration depth as a fraction of its ambient pressure value,

\begin{eqnarray}
\lambda_{L0}(P) \equiv \frac{\lambda_L(0, P)}{\lambda_L(0, 0)} = e^{\frac{KP}{2}}.
\end{eqnarray}

\noindent  When deriving this, we assume that the quasiparticle density $n$ remains the same for the all pressure values. Similarly, we provide the zero-temperature coherence length as a fraction of its ambient pressure value,

\begin{eqnarray}
\xi_{0}(P) \equiv \frac{\xi(0, P)}{\xi(0, 0)} = A e^{-2KP},
\end{eqnarray}

\noindent where we defined a constant $A = 4/[(g_0(\epsilon_F) v] e^{-2/[g_0(\epsilon_F) v]}$. Notice the exponential term in the definition of $A$, this is because we assume that the ambient pressure superhydrides are weak coupling superconductor. The coherence length and the Landau penetration depth can be used to find the pressure dependent Ginzburg-Landau parameter,

\begin{eqnarray}
\kappa_0(P) \equiv \frac{\kappa(0, P)}{\kappa(0, 0)} = \frac{1}{A}e^{\frac{5KP}{2}}.
\end{eqnarray}

\noindent As opposed to the other superconductors, we argue that the coherence length and the Landau penetration depth cannot be considered as relevant length scales for the superhydrides under pressure. This is due to the fact that they have strong pressure dependence as shown above. Thus, the classification of type-I versus type-II may not be appropriate for superhydrides, unless one specifies the pressure.

\section{V. Enhancement of the density of states}

In this section, we justify the validity of our theory by comparing the density of states. First, we extract the experimental density of states from the experimental measurements for the H$_3$S system. The lower critical field and the upper critical field within the BCS theory are given by~\cite{tinkham96},

\begin{eqnarray}
H_{C1}(T, P) = \frac{\phi_0}{4 \pi \lambda^2(T, P)} \ln [\kappa(T, P)],
\end{eqnarray}

\noindent and

\begin{eqnarray}
H_{C2}(T, P) = \frac{\phi_0}{2 \pi \xi^2(T, P)},
\end{eqnarray}

\noindent where $\phi_0 = hc/2e$ is the flux quantum. These two can be combined into a single equation to determine the Landau Ginzburg parameter using the lower and upper critical fields,

\begin{eqnarray}
\kappa^2(T, P) = \frac{H_{C2}(T, P)}{2 H_{C1}(T, P)} \ln[\kappa(T, P)].
\end{eqnarray}

\noindent Once the Landau Ginzburg parameter is known, the thermodynamic critical field,

\begin{eqnarray}
H_{C}(T, P) = \frac{\phi_0}{2\sqrt{2} \lambda_L(T, P) \xi(T, P)},
\end{eqnarray}

\noindent can be determined by,

\begin{eqnarray}
H_{C}(T, P) = \frac{H_{C2}(T, P)}{\sqrt{2} \kappa(T, P)}.
\end{eqnarray}

\noindent The zero temperature limit of the thermodynamic critical field is related to the density of states and the zero temperature gap function,

\begin{eqnarray}
H_{C}(0, P) = \sqrt{2 \pi g(\epsilon_F)} \Delta_0.
\end{eqnarray}

\noindent The pressure dependent superconducting gap $\Delta_0$ is related to the pressure dependent critical temperature, $\Delta_0 = 2 k_BT_C$, note the factor $2$ on the right-hand side in our theory as opposed to the factor of $1.763$ in standard weak coupling BCS approximation. Finally, the experimental density of states at a given pressure can be determined by using the experimental determination of thermodynamic critical field and the critical temperature,

\begin{eqnarray}
g(\epsilon_F) = \frac{H^2_C(0, P)}{8 \pi (k_BT_C)^2}.
\end{eqnarray}

\noindent Using the magnetization measurements, the lower critical field for the H$_3$S system has been extracted to be $H_{C1}(0) = 0.03 T$ by Drozdov \emph{et al}~\cite{SCEX1}. However, using the sample geometry of the NRS experiment~\cite{nmr}, Hirsch \emph{et al}~\cite{hirsch3} argued that $H_{C1} > 2.5 T$. Using this lower bound for the lower critical field and the experimental value for the upper critical filed $H_{C2} = 70 T$, Eq. (26) gives us $\kappa = 4.6$ for the H$_3$S system. Equation (28) then yields $H_C(0) = 10.8 T$. We then use the Eq. (30) to find the density of states for both spins~\cite{hirsch3}:

\begin{eqnarray}
g(\epsilon_F) = 1.053/eV {\AA}^3
\end{eqnarray}

\noindent This density of states is about 28 times larger than that of the ambient pressure sulfer hydride , $ g_0(\epsilon_F) = 0.038/eV{\AA}^3$~\cite{dos}. Using the $K = 0.021/GPa$, the density of states of the H$_3$S system at pressure $P = 155 GPa$, we find, $g_(\epsilon_F) =  g_0(\epsilon_F)e^{KP} \equiv 0.911/eV{\AA}^3$. This excellent agreement justifies our massive electron concept for the superhydrides under pressure.

\section{VI. Resistive broadening at the superconducting transition}

Type-II superconductors show a broadening in resistivity at the superconducting transition in the presence of an applied magnetic field. Below the critical temperature, when the applied magnetic field is smaller than the upper critical field, but larger than the lower critical field, the material enters into the mixed phase. In the mixed phase, the magnetic field penetrates into the material as flux quantum. The flux bundles appears as vortices with normal conducting core forced by the diverging superfluid velocity. These vortices interact through repulsive forces, mediated by the vortex currents, but stay together due to the magnetic pressure.

In the mixed phase, the circulating current causes the motion of the vortices. This motion causes the flux-flow resistivity which broadens the superconducting transition. The resistivity, caused by the dissipation, originates from the normal core current in the vortex and the supercurrent around it~\cite{Caroli64}. The vortex motion creates a disturbance to the supercurrent around the vortex, which results the creation of an electric field distribution~\cite{bardeen65}. To have the continuity of the electric field, a normal current circulates within the core of the vortex. This normal core current creates the first dissipation. The electric field, which is perpendicular to both vortex direction and the vortex velocity, is created due to the motion of the vortices in a magnetic field~\cite{kim69, josph65}. The second source of dissipation is created by this electric field outside the vortex core~\cite{tinkham96}. It has been shown that both of these dissipation have a similar order of magnitude~\cite{tinkham96, bardeen65}.

When a vortex is in motion, two forces can act on the vortex, the Lorentz force and the frictional force. The Lorentz force on a vortex includes both a Lorentz-like force caused by the magnetic field pressure gradient in an external current~\cite{tinkham64}, and the Magnus contribution caused by the relative motion between the vortex and the supercurrent~\cite{deGennes64, Nozieres66, Brandt95}. The Lorentz force is the only external force acting on a vortex in a clean system. However, the materials always have disorder and defects causing a frictional force on a moving vortex. This frictional force is important  in  restoring  the  vortex  motion  disturbed  by the dissipation in the vortex core~\cite{Kopnin76}. In a clean enough system, the vortex flow gives rise to a flux flow resistance due to these forces acting on a vortex. Using the condition for the dynamical equilibrium where the frictional force is equal to the Lorentz force, the flow resistivity has been derived~\cite{Stranad64, thesis},

\begin{eqnarray}
\rho(T, P) = \frac{2 \pi \xi^2(T, P) \rho_n(T, P) B}{\phi_0},
\end{eqnarray}

\noindent where $B$ is the applied magnetic field and $\rho_n(T, P) = m^\ast/(ne^2 \tau)$ is the normal-state resistivity with $\tau$ being the relaxation time. Taking the zero-temperature limits, the flux flow resistivity as a fraction of its ambient pressure value,

\begin{eqnarray}
\rho_{0}(P) \equiv \frac{\rho(0, P)}{\rho(0, 0)} = A^2 e^{-3KP}.
\end{eqnarray}

\noindent Note the exponentially decaying factor $e^{-3KP}$ for the H$_3$S and $C-S-H$ systems at their highest critical temperatures, $7.2 \times 10^{-5}$ and $3.5 \times 10^{-3}$, respectively. These are almost 4-orders of magnitude smaller and 3-orders of magnitude smaller than that of the ambient pressure resistivities, respectively. This low resistivity gives a larger conductivity for the supercurrent at the transition, therefore the resistivity broadening is very small as evident by the experiments.

\section{VII. Conclusions}

We proposed a concept of a massive electron scheme to explain the non-standard properties of high-temperature superhydride superconductors. We showed that the effective mass of the electron-quasiparticles exponentially increases with applied pressure and agrees with experimental critical temperatures. Our investigation showed that the superhydrides are strongly interacting-conventional BCS superconductors at high pressure due to the large density of states. The estimated density of states and conductivity at the transition, in the presence of a magnetic field, are consistent with the experimental observations. We showed that the coherence length, the London penetration depth, and the Landau–Ginsburg parameter all have strong pressure dependence, hence the traditional categorization of type-I versus type-II superconductors is not applicable to superhydrides. Further, we showed that the conductivity at the superconducting transition in the presence of magnetic field is 3-5 orders of magnitude larger than that of other superconductors. Therefore, the superconducting transition width is very narrow, similar to type-I superconductors as seen in experiments. This larger conductivity is due to the strong pressure dependence on the coherence length. In addition to H$_3$S and C-S-H systems, the LaH$_{10}$ system also shows near room temperature superconductivity under pressure~\cite{exp2}. Even though we have not compared LaH$_{10}$ data with our theory, we anticipate our theory is applicable to this system also.

\section{VIII.  ACKNOWLEDGMENTS}

We are grateful to Dr. Ranga Dias and his collaborators for sharing their experimental data with us. We further acknowledge valuable communications with Dr. Dias.

\end{document}